\documentclass[a4paper,fleqn,12pt]{cas-sc}
\usepackage{soul,xcolor}
\sethlcolor{cyan} 

\usepackage[numbers]{natbib}
\usepackage{amsmath}
\usepackage{longtable}
\usepackage{braket}

\ExplSyntaxOn
\cs_set:Npn \__first_footerline: {}
\ExplSyntaxOff

\begin{document}
\let\WriteBookmarks\relax
\def\floatpagepagefraction{1}
\def\textpagefraction{.001}

\shorttitle{Spectroscopy of Ho$^{3+}$ in K$_2$YF$_5$ }    

\shortauthors{P.\ Chanprakhon et.\ al.}  

\title [mode = title]{Infrared Absorption and Laser Spectroscopy of Ho$^{3+}$ Doped K$_2$YF$_5$ Microparticles}  

\author[1,2]{Pakwan Chanprakhon}[orcid=0000-0002-0343-8732]
\credit{Investigation, Formal analysis, Writing - original draft}
\author[1,2]{Michael F. Reid}[orcid=0000-0002-2984-9951]
\cormark[1]
\credit{Conceptualization, Formal analysis, Software, Investigation, Supervision, Writing - review and editing}
\ead{mike.reid@canterbury.ac.nz}
\author[1,2]{Jon-Paul R. Wells}[orcid=0000-0002-8421-6604]
\cormark[1]
\credit{Supervision, Formal analysis, Writing - review and editing, Visualization, Resources, Funding Acquisition, Project Administration}
\ead{jon-paul.wells@canterbury.ac.nz}

\affiliation[1]{organization={School of Physical and Chemical Sciences, University of Canterbury},
            addressline={PB4800, Christchurch 8140, New Zealand}}
          
\affiliation[2]{organization={Dodd-Walls Centre for Photonic and Quantum Technologies},
            addressline={New Zealand}}
        
\cortext[1]{Corresponding authors}

\fntext[1]{}


\begin{abstract}
High-resolution absorption and laser spectroscopy are used to  determine electronic energy levels for  Ho$^{3+}$ ions in K$_2$YF$_5$ microparticles. A total of 72 crystal-field energy levels, distributed among 8 multiplets, are assigned. This optical data is used for crystal-field modelling of the electronic structure of Ho$^{3+}$ in K$_2$YF$_5$. Partially-resolved hyperfine splittings are accurately reproduced by the model. The temperature dependence of the fluorescent lifetime of the $^5$F$_5$ multiplet is measured and the temperature dependence of the non-radiative relaxation is modelled  by a five-phonon process. Preliminary measurements of infra-red to visible upconversion in microparticles co-doped with Ho$^{3+}$ and Yb$^{3+}$ is reported. 
%
\end{abstract}

  



\maketitle

\section{Introduction}

It is a pleasure to dedicate this paper to Professor Hans G{\"u}del, who has made outstanding contributions to the spectroscopy of materials doped with transition-metal and lanthanide ions, including seminal studies on upconversion fluorescence. The investigation reported here is motivated, in part, by his work on spectroscopy, radiative and non-radiative relaxation, and upconversion in lanthanide-doped halide crystals. 

Insulating, dielectric nano-crystals, optically activated by transition series ions, are of considerable current interest for applications in biomedical imaging \cite{Dong2015, Bouzigues2011}, optical-based sensing such as nano-thermometry \cite{Bouzigues2011, Himmelsto, Pratik2021}, and quantum technologies such as cryogenic quantum memories \cite{Thiel2011, Martin2023}.  Lanthanide ion doped potassium yttrium pentafluoride crystals (K$_2$YF$_5$) have been less commonly investigated although it is a mechanically hard, thermally stable and optically transparent material over a wide range; offering a single substitutional site for occupancy by lanthanide ions. The K$_{2}$YF$_5$ crystal is a member of the orthorhombic system having space group Pna21 and unit cell dimensions $a=10.820$\,\AA, $b=6.613$\,\AA, $c=7.249$\,\AA. In the
K$_2$YF$_5$ structure, each Y$^{3+}$ ion is surrounded by seven fluoride ions, giving C$_{\rm s}$ point group symmetry. The YF$_7$ polyhedra
form chains parallel to the c-axis \cite{karbowiak2012energy,zverev2011electron}. The Y$^{3+}${}- Y$^{3+}$ intra-chain distance is approximately 3.7\,\AA,
whereas the shortest inter-chain separation is \~{}5\,\AA ~\cite{loiko2016up}. Lanthanide ions substitute for the Y$^{3+}$ cation in the K$_{2}$YF$_{5}$ host material and therefore occupy sites of very low point group symmetry.

In this study, we undertake detailed spectroscopic measurements for K$_2$YF$_5$ microparticles doped with Ho$^{3+}$ ions, which has not been previously reported in the literature. The microcrystals were prepared using a standard hydrothermal synthesis method and their morphology and structure were determined by SEM and X-ray diffraction. From a combination of high resolution absorption and laser excited fluorescence measurements, a total of 72 experimental crystal-field levels have been determined for multiplets whose transitions span the infrared through to the visible regions. As is commonly the case for Ho$^{3+}$ doped materials (see for example \cite{Mujaji1992,  Wells2004, Popova2017, Smith2023, Mothkuri2021, Mothkuri2024}),
the ground state consists of two closely-spaced levels. Wavefunctions derived from a crystal-field calculation are used to estimate the hyperfine structure generated by the pseudo-quadrupole interaction between these nearby singlet levels, which accounts for the splittings and asymmetry observed in absorption spectra.
Temperature-dependent measurements of the lifetime of the $^5$F$_5$ multiplet are presented. A preliminary study of upconversion fluorescence for K$_{2}$YF$_{5}$:Yb$^{3+}$/Ho$^{3+}$ using excitation at 980 nm is reported.

\section{Materials and methods}

\subsection{Synthesis of K$_2$YF$_5$:Ho$^{3+}$ microparticles}

The K$_2$YF$_5$ samples were synthesized using a hydrothermal technique \cite{Bian2019}. A solution was prepared by dissolving 56.25 mmol of KOH in 7.5 mL and 120 mmol of KF in 10 mL of deionized water. After the initial mixing, 25 ml of ethanol and 20 mL of oleic acid were added and mixed for 10 minutes. Finally, 3 mmol of Y(NO$_3$)$_3$$\cdot$6H$_2$O + $x$\%Re(NO$_3$)$_3$, was added to the solution and stirred until it was homogeneous, then transferred to a 100-mL autoclave and put in the oven at 220$^\circ$C for 24 hours. The microparticles were separated through centrifugation and washed 3 times with ethanol, deionized water, and ethanol sequentially. Singly-doped samples with 0.25\%, 0.5\%, 1\%, and 2\% Ho$^{3+}$, as well as co-doped samples with 20\% Yb$^{3+}$ and 2\% Ho$^{3+}$, were prepared in this fashion.

\subsection{Characterisation}

X-ray diffraction (XRD) data was collected using a RIGAKU 3 kW SmartLab X-ray diffraction spectrometer with a Cu-K$\alpha$ radiation source. The measurements were conducted at 40 kV and 30 mA, scanning from 5 to 65 degrees. The morphology was measured by the JOEL 7000F Scanning Electron Microscope. The size distribution of the samples was determined using the ImageJ program to track the size of the microparticles.

\subsection{Absorption and fluorescence}

Absorption spectra were measured using a Bruker Vertex 80 FTIR having a spectral range of 400-25,000 cm$^{-1}$ at a maximum spectral resolution of 0.075 cm$^{-1}$. The microparticle samples were compressed into a thin pellet and mounted onto the cold finger of a Janis closed-cycle helium compressor which cooled the samples to a nominal temperature of 7~K. A combination of InGaAs and silicon photodetectors were used. 

For laser spectroscopy the samples were excited using a PTI wavelength tunable, pulsed dye laser employing either Coumarin 460, 481, and Rhodamine 640 laser dyes as appropriate. Temperature dependent studies were obtained using a Janis closed cycle cryostat with a resistive heater coupled to a Lakeshore temperature controller. An iHR550 single monochromator and Peltier cooled Hamamatsu R2257P and H10330C photomultiplier tubes were used to record the fluorescence spectra.

Upconversion measurements were made using a temperature-controlled InGaAs laser diode operating at a centre wavelength of 980\,nm. Low-resolution fluorescence was recorded with an Ocean Optics (USB2000) mini-spectrometer.
The excitation power was measured using the Coherent FieldMaxII-TOP power meter.

\subsection{Crystal-field calculations}

The 4f$^{11}$ configuration appropriate to Ho$^{3+}$ can be described using a parametrized Hamiltonian \cite{wybourne1965spectroscopic,Carnall1989,reid2016theory}:
\begin{equation}
    H = H_\text{FI} + H_\text{CF} + H_\text{HF}
\end{equation}
where $H_\text{FI}$ is the free ion contribution,  $H_\text{CF}$ is the crystal-field contribution, and $H_\text{HF}$ is the hyperfine interaction. Further details of the free-ion Hamiltonian can be found in the above references. The crystal-field Hamiltonian may be written as
\begin{equation}
\label{eq:cf}
    H_\text{CF} = \sum_{k,q} B_{q}^{k}C_{q}^{(k)}, 
\end{equation}
where the  $B_{q}^{k}$ are crystal-field parameters and the $C_{q}^{(k)}$ are Racah spherical tensor operators for the 4f$^N$ configuration. For 4f electrons $k$ is restricted to 2, 4, and 6. In C$_{\rm s}$ symmetry, only parameters with even $k$ are non-zero and the parameters with $q \neq 0$ are imaginary. We base our calculation on our previous work on K$_2$YF$_5$:Er$^{3+}$ \cite{Solanki2024}.  Er$^{3+}$  has an odd number of 4f electrons, so the electronic states are doubly degenerate. Ho$^{3+}$ has an even number of electrons, so all electronic states are singlets. 

Trivalent holmium has one stable isotope, $^{165}$Ho, with a nuclear spin of 7/2, so each electronic singlet will split into four doubly-degenerate states. Owing to its large nuclear magnetic moment, hyperfine structure is commonly observable by conventional spectroscopy. Details of our computational approach are given in Refs.\ \cite{Wells2004, Smith2023, Mothkuri2021}.

\section{Results and discussion}
\subsection{Phase, morphology, and composition}

\begin{figure}[tb!]
\centering
 \includegraphics[width=0.5\textwidth]{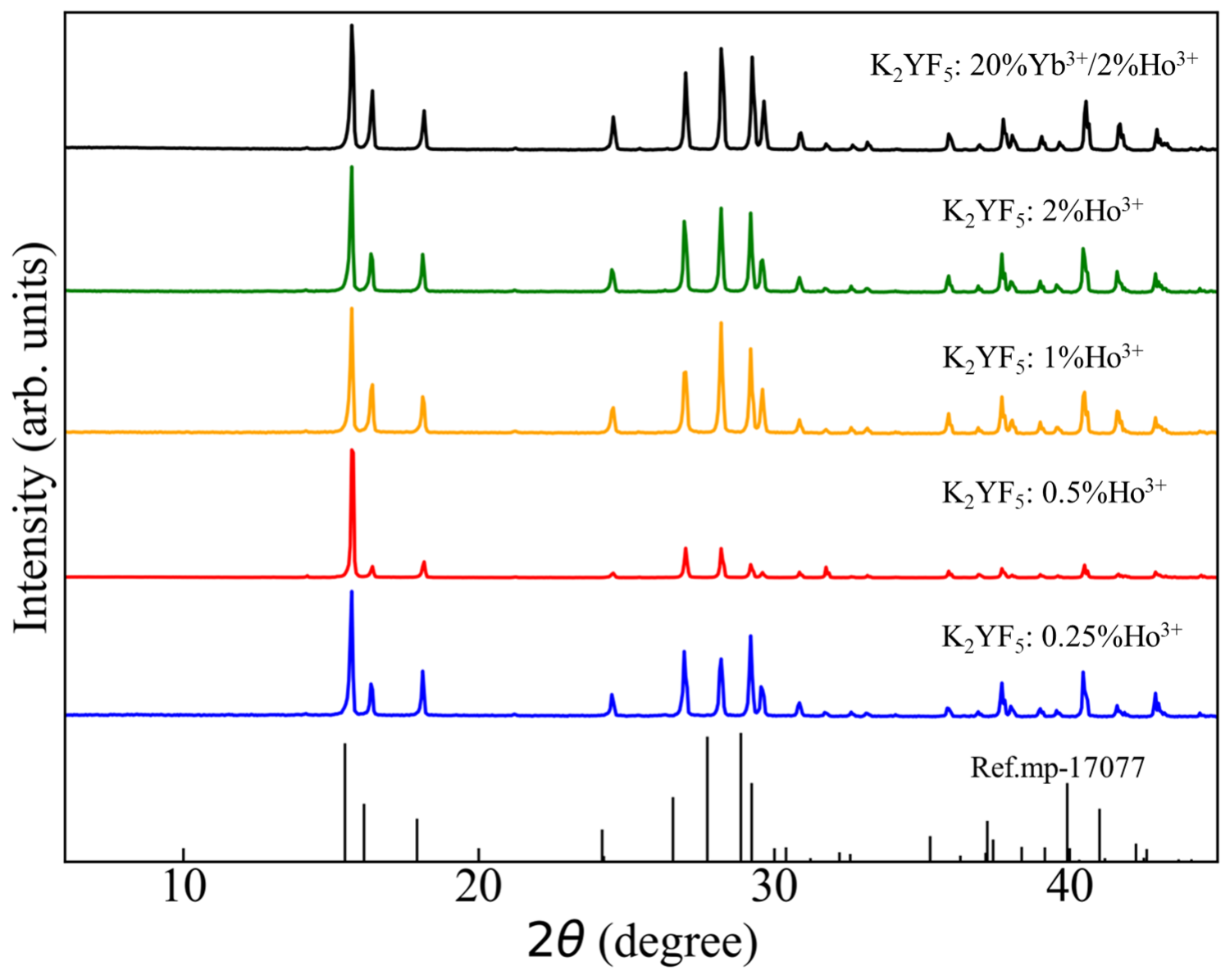} 
\caption{ \label{fig:xrd}
XRD patterns of K$_2$YF$_2$ microparticles with lanthanide ion dopant concentrations between 0.25 and 22 molar percent}
\end{figure}

\begin{figure}[tb!]
\centering
 \includegraphics[width=\textwidth]{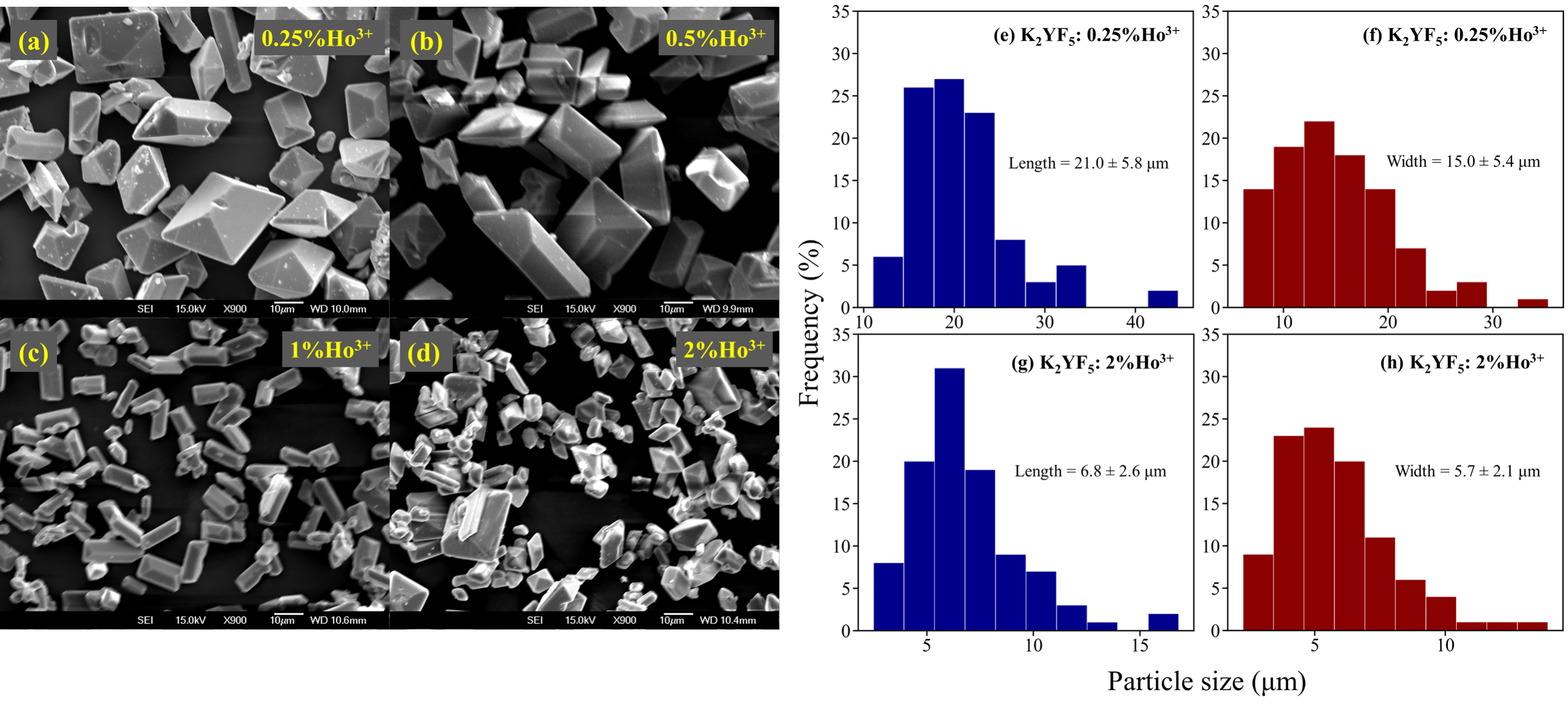} 
\caption{ \label{fig:sem}
(a-d) SEM images and (e-h) particle size distributions for K$_2$YF$_5$ microparticles with Ho$^{3+}$ concentrations ranging from 0.25\% to 2\%.}
\end{figure}

The powder XRD patterns of the as prepared K$_2$YF$_5$ samples are shown in Figure \ref{fig:xrd} and compared to standard reference data (mp-17077). The diffraction patterns of five different samples are in good agreement with the reference data for orthorhombic K$_{2}$YF$_{5}$ having a Pna21 space group. No additional phase is evident. Small shifts in the peaks are apparent as the dopant concentration is increased. The SEM results in Figure \ref{fig:sem} show that the microparticles have an octahedral shape, and the particle size decreases with higher dopant concentration. The particle size distribution indicates that samples doped with 0.25\% Ho$^{3+}$ have an average length of 21 $\pm$ 5.8 $\mu$m which decreases to 6.8 $\pm$ 2.6 $\mu$m for the sample doped with 2\% Ho$^{3+}$.

\clearpage
\subsection{Absorption and laser excited fluorescence measurements}

\begin{figure}[tb!]
\centering
  \includegraphics[width=0.9\textwidth]{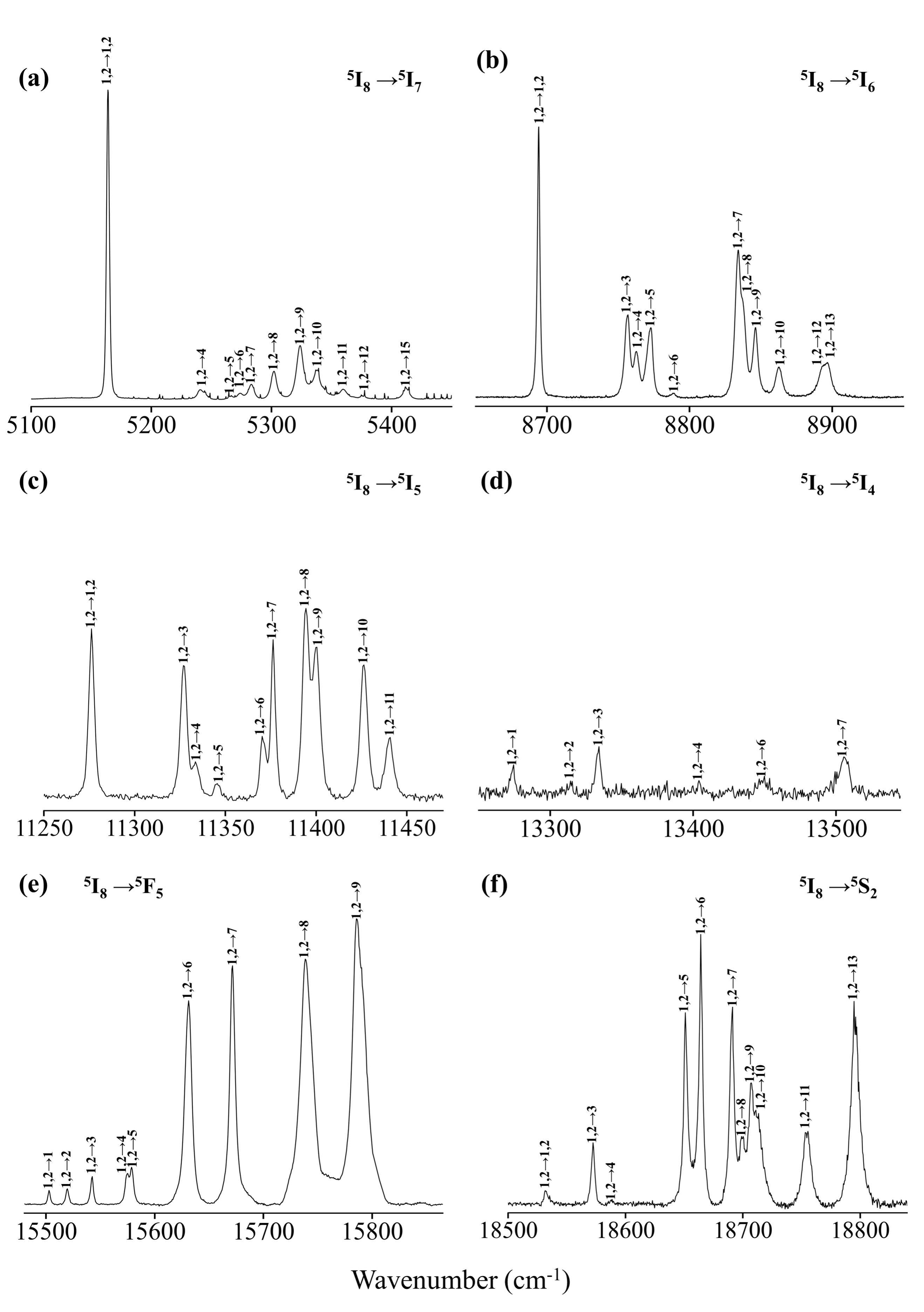} 
  \caption{\label{fig:absorption}
  The absorption spectra of K$_2$YF$_5$: 2\% Ho$^{3+}$ microparticles at 7 K, showing the transitions for
(a) $^5$I$_8$ $\rightarrow$  $^5$I$_7$;
(b) $^5$I$_8$ $\rightarrow$  $^5$I$_6$;
(c) $^5$I$_8$ $\rightarrow$  $^5$I$_5$;
(d) $^5$I$_8$ $\rightarrow$  $^5$I$_4$;
(e) $^5$I$_8$ $\rightarrow$  $^5$F$_5$;
(f) $^5$I$_8$ $\rightarrow$  $^5$S$_2$.}
\end{figure}

\begin{figure}[tb!]
\centering
 \includegraphics[width=0.9\textwidth]{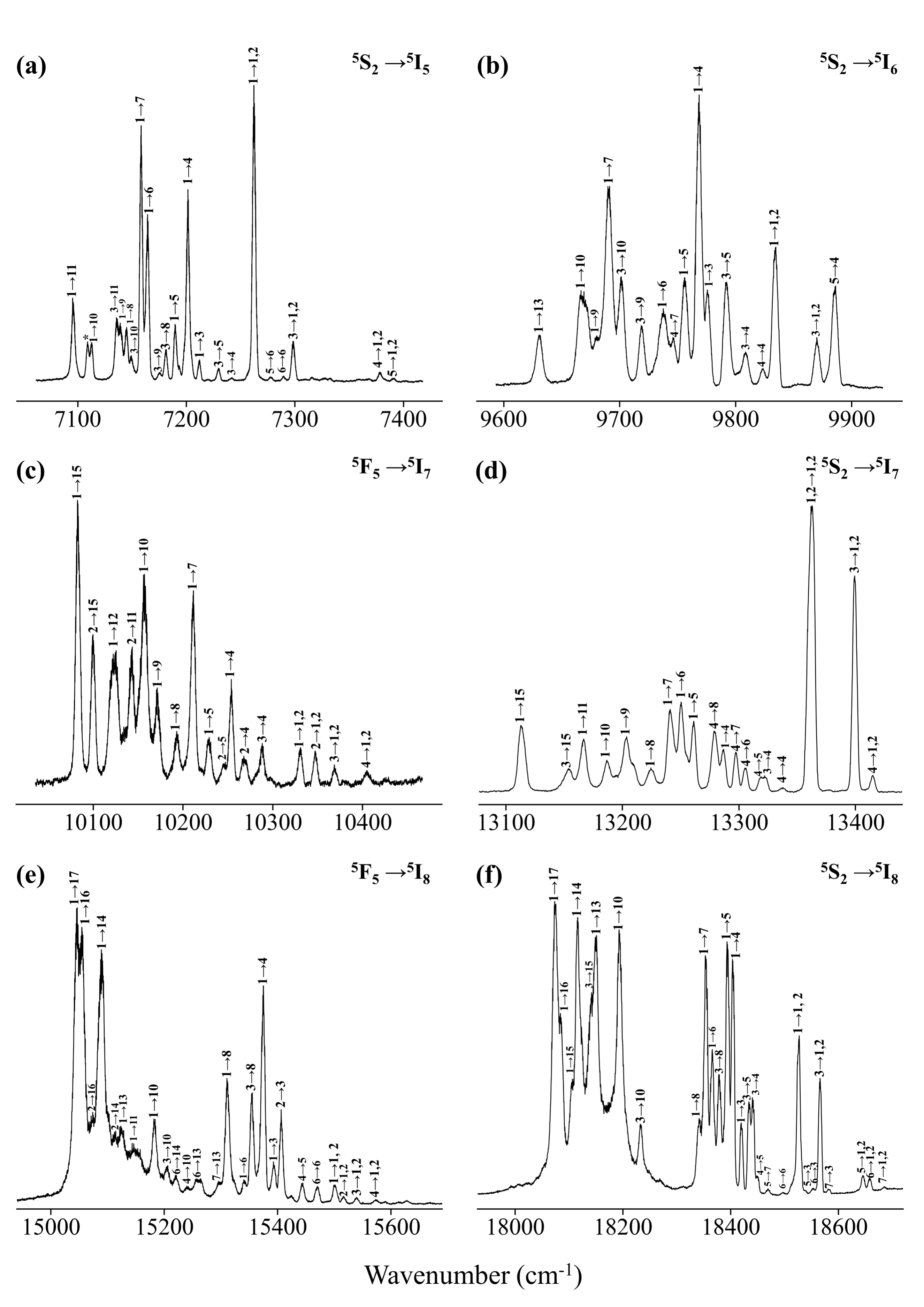} 
\caption{ \label{fig:emission}
The fluorescence spectra of K$_2$YF$_5$:0.5\% Ho$^{3+}$ microparticles measured at 10 K,
showing the
transitions for
(a) $^5$S$_2$ $\rightarrow$  $^5$I$_5$;
(b) $^5$S$_2$ $\rightarrow$  $^5$I$_6$;
(c) $^5$F$_5$ $\rightarrow$  $^5$I$_7$;
(d) $^5$S$_2$ $\rightarrow$  $^5$I$_7$;
(e) $^5$F$_5$ $\rightarrow$  $^5$I$_8$;
(f) $^5$S$_2$ $\rightarrow$  $^5$I$_8$.
* denotes an unassigned spectral feature. 
}
\end{figure}

The absorption spectra of K$_2$YF$_5$ microparticles doped with 2\% Ho$^{3+}$ were collected using a pelletized sample cooled to a nominal temperature of 7 K. A sample doped with 2\% Ho$^{3+}$ was chosen to give good absorption depth. Figure \ref{fig:absorption} presents the measured absorption spectra for transitions from the ground state of $^5$I$_8$ to seven different excited multiplets in the NIR and visible regions. Fifty eight crystal-field levels could be assigned directly from absorption measurements.
For the lowest energy transition to the $^{5}$I$_{7}$ multiplet (at 5163.9 cm$^{-1}$) two peaks are observed, with a separation of 0.5 cm$^{-1}$. This suggests the existence of pairs of singlet states (in C$_{\rm s}$ symmetry all states are singlets), which will be discussed in detail below. The sharp features in Figure \ref{fig:absorption}(a) are residual atmospheric water absorption lines present due to incomplete purging of the beam path with N$_{2}$ gas.

Laser excited fluorescence spectra for K$_2$YF$_5$:0.5 \%Ho$^{3+}$ are shown in Figure \ref{fig:emission}, for fluorescence emanating from the ${^5}$S$_{2}$ and $^{5}$F$_{5}$ multiplets. The spectra shown were recorded for samples cooled to a nominal temperature of 10~K and for excitation at 20,742 cm$^{-1}$ (482.1 nm), corresponding to one of the $^5$I$_8$ $\rightarrow$ $^5$F$_3$ transitions. A lower dopant concentration was used for fluorescence experiments to circumvent any deleterious fluorescence quenching effects at high dopant concentration \cite{Gomes1996}. Fluorescence to the $^{5}$I$_{8}$  ground multiplet allowed for the assignment of 14 crystal field levels of that multiplet whilst fluorescence to the excited multiplets of the $^{5}$I term confirmed assignments based on absorption measurements. It is noticeable through the fluorescence spectra that transitions from thermally populated excited states are also present. This occurs due to the difficulty in getting good thermal contact across the entire pellet with the copper sample mount, however the additional transitions present are helpful in making assignments. All of the experimentally assigned energy levels are given in Table~\ref{tab:levels}. 

\clearpage

\begin{longtable}{ccr@{\hskip 10mm}r}
\caption{ 
  Experimental and calculated electronic energy levels for Ho$^{3+}$ doped in K$_2$YF$_5$ (in cm$^{-1}$). The experimental uncertainties are 1 cm$^{-1}$ for the $^5$I$_8$ multiplet and 0.1 cm$^{-1}$ for the other multiplets since those were determined from absorption. The alphanumeric labels follow the standard Dieke convention \cite{Dieke}. $^\ddag$The Z$_{1}$-Z$_{2}$ splitting was inferred from a fit to the absorption lineshape for Z$_1$, Z$_2$ $\rightarrow$ Y$_1$,Y$_2$, including a pseudo-quadrupole hyperfine interaction between the states. }
\label{tab:levels}\\


\hline
Multiplet         &State             & Measured            &Fit \\
\hline

$^5$I$_8$       &Z$_1$             &            0             &          0          \\
                &Z$_2$             &           1.35$^\ddag$       &         1.35         \\
                &Z$_3$             &           106            &         110.0       \\
                &Z$_4$             &           122            &         115.3       \\
                &Z$_5$             &           132            &         136.7       \\
                &Z$_6$             &           160            &         155.2       \\
                &Z$_7$             &           172            &         186.6       \\
                &Z$_8$             &           184            &         188.9       \\           
                &Z$_9$             &             -            &         309.6       \\           
                &Z$_{10}$          &           332            &         328.0       \\
                &Z$_{11}$          &           352            &         361.7       \\
                &Z$_{12}$          &            -             &         369.8       \\
                &Z$_{13}$          &           376            &         378.6       \\
                &Z$_{14}$          &           409            &         390.1       \\
                &Z$_{15}$          &           419            &         400.8       \\
                &Z$_{16}$          &           442            &         436.3       \\       
                &Z$_{17}$          &           451            &         439.9       \\ \\
                  
$^5$I$_7$       &Y$_1$             &          5163.9          &         5144.8      \\
                &Y$_2$             &            -             &         5145.2      \\
                &Y$_3$             &            -             &         5223.4      \\
                &Y$_4$             &          5242.1          &         5229.4      \\
                &Y$_5$             &          5266.3          &         5261.8      \\
                &Y$_6$             &          5274.1          &         5263.5      \\
                &Y$_7$             &          5282.3          &         5278.9      \\
                &Y$_8$             &          5302.1          &         5297.4      \\          
                &Y$_9$             &          5323.9          &         5312.9      \\          
                &Y$_{10}$          &          5337.0          &         5357.1      \\
                &Y$_{11}$          &          5359.8          &         5365.5      \\ 
                &Y$_{12}$          &          5376.2          &         5383.4      \\
                &Y$_{13}$          &            -             &         5395.4      \\
                &Y$_{14}$          &            -             &         5401.8      \\          
                &Y$_{15}$          &          5412.3          &         5410.8      \\ \\
                
$^5$I$_6$       &A$_1$             &          8694.2          &         8676.6      \\
                &A$_2$             &            -             &         8678.9      \\
                &A$_3$             &          8756.3          &         8744.8      \\
                &A$_4$             &          8762.8          &         8746.2      \\
                &A$_5$             &          8772.4          &         8765.2      \\
                &A$_6$             &          8788.7          &         8775.8      \\ 
                &A$_7$             &          8834.2          &         8784.7      \\
                &A$_8$             &          8836.7          &         8847.1      \\
                &A$_9$             &          8846.2          &         8855.6      \\
                &A$_{10}$          &          8862.6          &         8881.8      \\
                &A$_{11}$          &            -             &         8882.6      \\             
                &A$_{12}$          &          8893.2          &         8890.9      \\              
                &A$_{13}$          &          8896.7          &         8892.5      \\ \\
                
$^5$I$_5$       &B$_1$             &         11276.3          &        11254.8      \\
                &B$_2$             &           -              &        11255.6      \\
                &B$_3$             &         11327.2          &        11316.9      \\
                &B$_4$             &         11333.3          &        11320.2      \\
                &B$_5$             &         11345.7          &        11342.0      \\
                &B$_6$             &         11370.8          &        11388.3      \\
                &B$_7$             &         11376.3          &        11389.4      \\          
                &B$_8$             &         11394.5          &        11405.8      \\
                &B$_9$             &         11400.0          &        11425.5      \\
                &B$_{10}$          &         11426.2          &        11426.8      \\  
                &B$_{11}$          &         11440.4          &        11438.8      \\ \\

$^4$I$_4$       &C$_1$             &         13273.4          &        13235.8      \\
                &C$_2$             &         13314.1          &        13266.9      \\
                &C$_3$             &         13333.8          &        13370.2      \\
                &C$_4$             &         13403.6          &        13400.8      \\
                &C$_5$             &            -             &        13412.9      \\
                &C$_6$             &         13449.1          &        13470.4      \\
                &C$_7$             &         13505.5          &        13520.2      \\  
                &C$_8$             &            -             &        13534.3      \\        
                &C$_9$             &            -             &        13643.5      \\ \\
                
$^5$F$_5$       &D$_1$             &         15502.3          &        15501.4      \\
                &D$_2$             &         15519.5          &        15504.6      \\
                &D$_3$             &         15542.3          &        15554.9      \\
                &D$_4$             &         15574.7          &        15560.8      \\
                &D$_5$             &         15578.8          &        15563.6      \\
                &D$_6$             &         15630.9          &        15673.9      \\
                &D$_7$             &         15671.5          &        15681.6      \\  
                &D$_8$             &         15739.4          &        15757.8      \\  
                &D$_9$             &         15788.3          &        15771.0      \\
                &D$_{10}$          &            -             &        15807.9      \\
                &D$_{11}$          &            -             &        15809.1      \\ \\
                
$^5$S$_2$       &E$_1$             &         18533.0          &        18534.0      \\
                &E$_2$             &            -             &        18538.9      \\
                &E$_3$             &         18572.2          &        18587.6      \\
                &E$_4$             &         18587.9          &        18590.3      \\
                &E$_5$             &         18650.9          &        18597.2      \\ \\

$^5$F$_4$       &E$_6$             &         18664.6          &        18655.5      \\
                &E$_7$             &         18691.1          &        18692.3      \\
                &E$_8$             &         18699.4          &        18694.3      \\
                &E$_9$             &         18709.9          &        18714.8      \\
                &E$_{10}$          &         18711.8          &        18730.5      \\
                &E$_{11}$          &         18754.5          &        18740.5      \\
                &E$_{12}$          &            -             &        18773.0      \\  
                &E$_{13}$          &         18796.0          &        18806.7      \\
                &E$_{14}$          &            -             &        18807.2      \\ \\
  
\hline
\end{longtable}
\clearpage

\begin{table}[b!]
\caption{ \label{tab:crystalfield}
  Free-ion, crystal-field, and hyperfine parameters for Ho$^{3+}$-doped K$_2$YF$_5$ (in cm$^{-1}$).
  Free-ion parameters that were fixed to the values obtained for Ho$^{3+}$ in LaF$_3$ \cite{Carnall1989} are not shown. 
  Parameters in square brackets were not varied. The crystal-field parameters were constrained so that the ratios of the parameters with $q \neq 0$ to $q=0$ were the same as for Er$^{3+}$ \cite{Solanki2024}. The magnetic and electric-quadrupole hyperfine parameters A and Q are those used in Ref.\ \cite{Wells2004}. 
}
\footnotesize
\renewcommand{\arraystretch}{1.3}
\begin{tabular}{cc@{\hskip 10mm}c@{\hskip 10mm}c@{\hskip 10mm}c}
\hline
Parameter       & Fit            \\
\hline                                                                
                                                                        
$E_\text{avg}$  &       48524        \\
$F^2$           &       94059        \\   
$F^4$           &       66353        \\   
$F^6$           &       51494        \\    
$\zeta$         &       2143         \\      
$B^2_0$         &       -399         \\ 
$B^2_2$         &       [-94-429i]     \\ 
$B^4_0$         &       -1250        \\ 
$B^4_2$         &       [-71-64i]     \\ 
$B^4_4$         &       [-1285-141i]   \\ 
$B^6_0$         &       281          \\ 
$B^6_2$         &       [140-89i]      \\ 
$B^6_4$         &       [-354-106i]    \\ 
$B^6_6$         &       [-200+194i]    \\ 
$A$             &       [0.037]   \\
$Q$             &       [0.06]    \\
\hline
\end{tabular}
\end{table}
  
\clearpage

\subsection{Crystal-field analysis}
In C$_{\rm s}$ symmetry, crystal-field parameters, $B^k_q$, with $k = 2$, $4$, and $6$ and $q$ even are non-zero and the parameters with $q \neq 0$ are imaginary. There are, therefore, 15 crystal-field parameters (counting the real and imaginary parts separately).  
In our calculations for Er$^{3+}$ in K$_2$YF$_5$ \cite{Solanki2024}, we made use of magnetic-splitting data  \cite{zverev2011electron}. This made it practical to fit all of the crystal-field parameters using the computational approach of Ref.\ \cite{horvath2019}. Magnetic-splitting data is not available for  Ho$^{3+}$-doped K$_2$YF$_5$ so in this case we only varied the crystal-field parameters with $q=0$ and restricted the $q \neq 0$ crystal-field parameters to the ratios obtained for Er$^{3+}$. Since the electronic structure of Ho$^{3+}$ only differs from Er$^{3+}$ by the removal of one 4f electron, we expect the parameters to be very similar. In our fit, the ratios of the Ho$^{3+}$ to Er$^{3+}$ parameters are  1.05 for $B^2_q$,  0.99 for $B^4_q$, and 1.04 for $B^6_q$. The free ion parameters listed in Table \ref{tab:crystalfield} were allowed to vary, while the other parameters were taken from Ho$^{3+}$ in LaF$_3$ \cite{Carnall1989}. For the fitting, we constrained the splitting between Z$_1$ and Z$_2$ to 1.35 cm$^{-1}$. This was essential to accurately reproduce the lineshape of the $^5$I$_8$ (Z$_1$, Z$_2$) $\rightarrow$ $^5$I$_7$ (Y$_1$,Y$_2$) transition discussed in the following section. The results of the crystal field fit is shown in Table \ref{tab:levels}. The fitting of these data yielded the parameters in Table \ref{tab:crystalfield}. The fitting has a standard deviation of 18\,cm$^{-1}$. This could be reduced by allowing more crystal-field parameters to vary, but in the absence of magnetic-splitting data, the validity of the parameters would be impossible to determine. Since the ratios of the crystal-field parameters were constrained, we cannot give realistic estimates of uncertainties.

\subsection{Hyperfine structure}

\begin{figure}[b!]
\centering
\includegraphics[width=0.5\textwidth]{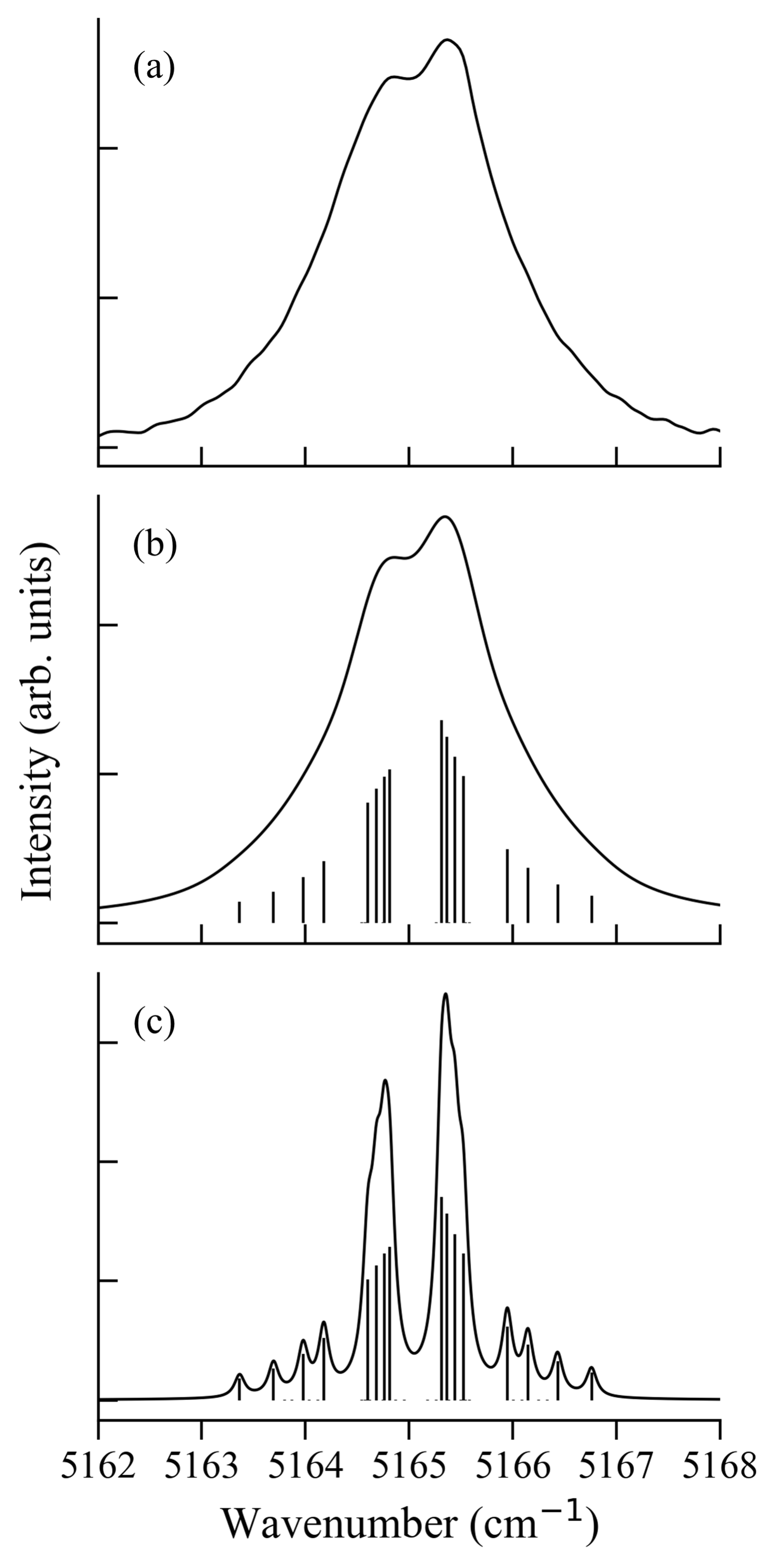}
\caption{\label{fig:hyperfine}
(a) The measured absorption spectrum of K$_2$YF$_5$0.25\% Ho$^{3+}$ showing the  $^5$I$_8$ (Z$_1$, Z$_2$) $\rightarrow$ $^5$I$_7$ (Y$_1$,Y$_2$) transition, measured at a nominal sample temperature of 7 K.  
(b) A simulation of the spectrum based on a Lorentzian lineshape for the transitions between hyperfine sub-levels assuming a linewidth 0.85 cm$^{-1}$. 
(c) A simulation of the spectrum based on a Lorentzian lineshape for the transitions between hyperfine sub-levels assuming a linewidth of 0.12 cm$^{-1}$. The simulations assumed a temperature of 10\,K. 
}
\end{figure}

Figure \ref{fig:hyperfine}(a) presents the absorption spectrum for the transition of $^5$I$_8$ (Z$_1$, Z$_2$) $\rightarrow$ $^5$I$_7$ (Y$_1$,Y$_2$) at a doping concentration of 0.25\% Ho$^{3+}$ and a nominal sample temperature of 7 K.
The hyperfine calculation was performed using the fitted crystal-field parameters, with the addition of hyperfine coupling between the 4f electrons and the nuclear spin taken from Ref.\ \cite{Wells2004}. In the crystal-field fit, which uses only the electronic Hamiltonian, the Z$_1$$-$Z$_2$ splitting was constrained to be 1.35\,cm$^{-1}$ and the Y$_1$$-$Y$_2$ splitting was calculated to be 0.40\,cm$^{-1}$. With the addition of the hyperfine Hamiltonian, each of these four states splits into four doublets.
The hyperfine interaction can not split isolated electronic singlets, as the diagonal matrix elements are zero, but there is a pseudo-quadrupolar interaction between closely-spaced singlets that gives an overall splitting of approximately 0.3\,cm$^{-1}$ for the four doublets from each of Z$_1$ and Z$_2$ and 0.5\,cm$^{-1}$ for the four doublets from each of Y$_1$ and Y$_2$.
This pattern is similar to Figure 2 of Ref.\ \cite{Mothkuri2021}. In that case the electronic states are further apart, so the hyperfine splitting is smaller.

Figure \ref{fig:hyperfine}(b) presents a simulation of the spectrum assuming a Lorentzian lineshape for the transitions between hyperfine sub-levels with a full-width half-maximum linewidth of 0.85 cm$^{-1}$ and a temperature of 10\,K, slightly warmer than the nominal sample temperature of 7 K. A Gaussian lineshape did not give satisfactory results.  The vertical lines in the Figure are calculated magnetic-dipole intensities for each transition, with $Z_1$ and $Z_2$ populations calculated from the Boltzman distribution. Note that there are only 16 transitions with significant intensity, as the nuclear-spin quantum number, $M_I$, is conserved. 
To optimise the simulation, we adjusted the Z$_1$$-$Z$_2$ splitting in the crystal-field fit, the linewidth, and the temperature. Thus, even though we could not resolve the hyperfine structure, it still provided valuable input to constrain the crystal-field fit. 

The line broadening makes the resolution of individual hyperfine transitions impossible. This broadening is expected in microcrystals with relatively high doping concentrations. In low-concentration single crystals, the structure should be resolvable.
Figure \ref{fig:hyperfine}(c) shows a simulation of the spectrum with the linewidths set to 0.12 cm$^{-1}$, which is commonly observed in low-concentration single crystals. In this case much of the structure should be resolvable.
This Figure may be compared with Figure 1 of Ref.\ \cite{Mothkuri2021} and Figure 6 of Ref.\ \cite{Wells2004}, which show similar transitions for Ho$^{3+}$ centres in Y$_2$SiO$_5$ and  CaF$_2$ respectively. We note that in the CaF$_2$ spectrum the Y$_1$ state is a doublet, whereas in the current case and in Y$_2$SiO$_5$  there are two closely-spaced  singlets.


\subsection{Fluorescence lifetimes}

\begin{figure}[b!]
\centering
 \includegraphics[width=0.5\textwidth]{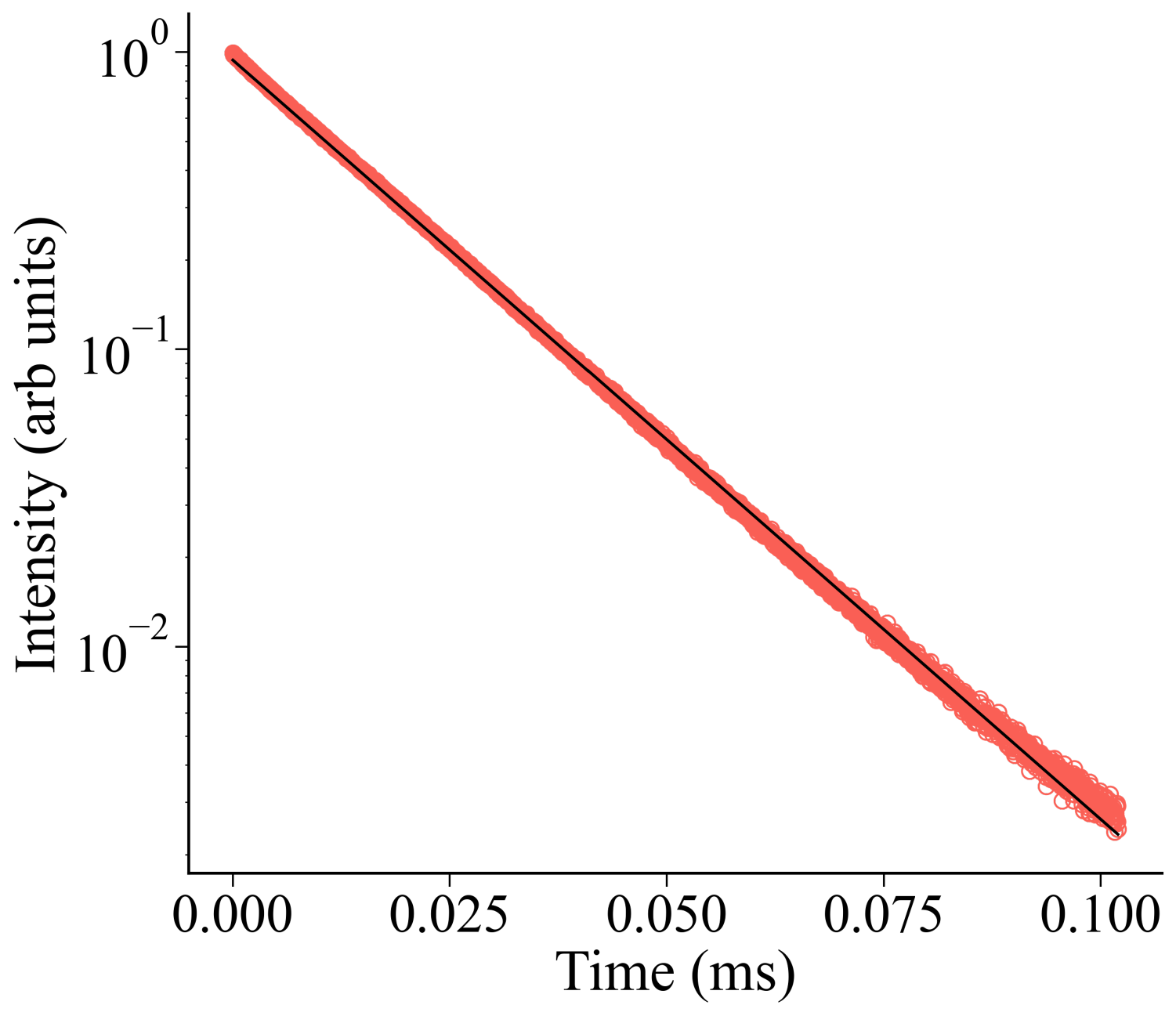} 
\caption{ \label{fig:5F5_lifetime}
The 10~K $^5$F$_5$ fluorescence transient. The solid line is a fit to a single exponential function.
}
\end{figure}

\begin{figure}[b!]
\centering
 \includegraphics[width=0.5\textwidth]{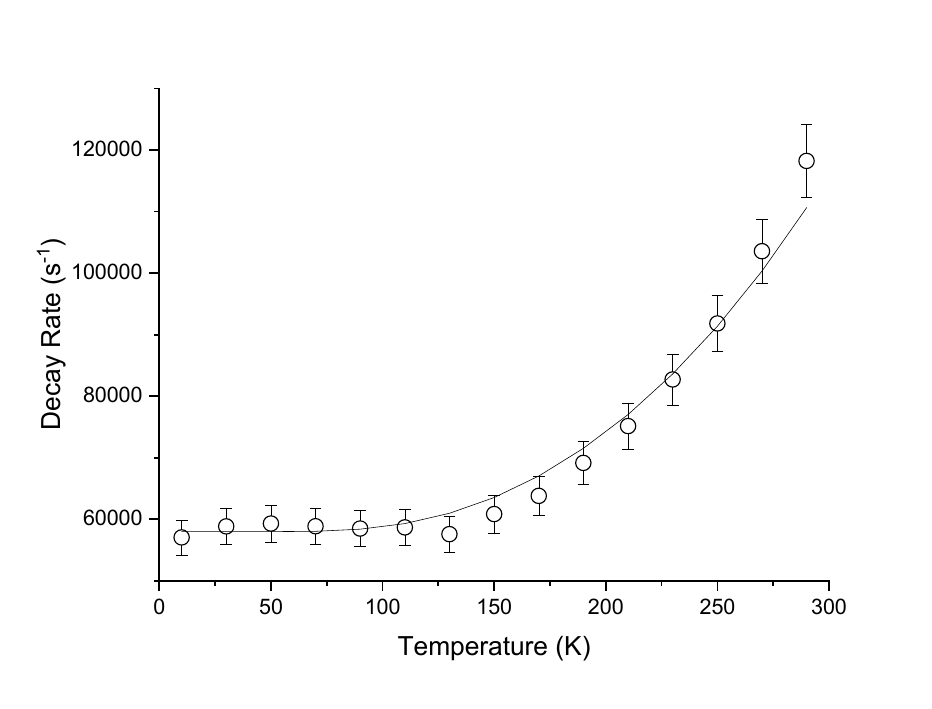} 
\caption{ \label{fig:5F5decay}
The temperature dependence of the $^5$F$_5$ fluorescence decay rate. The solid line is a fit to equation \ref{eq:decay}, as described in the text.
}
\end{figure}

The fluorescence lifetimes of the $^{5}$S$_{2}$ and $^{5}$F$_{5}$ multiplets for the 2\% Ho$^{3+}$ samples were measured. Figure \ref{fig:5F5_lifetime} shows the transient for a nominal sample temperature of 10~K. The $^{5}$S$_{2}$ multiplet was excited directly with fluorescence monitored at the E$_{1}\rightarrow$Z$_{17}$ transition (at 18082 cm$^{-1}$) whilst the $^{5}$F$_{5}$ multiplet was also excited directly with fluorescence monitored at the D$_{1}\rightarrow$Z$_{17}$ transition (at 15051 cm$^{-1}$). Both measured transients could be well approximated using a single exponential function yielding lifetimes at 10\,K  of 741\,$\mu$s for the $^{5}$S$_{2}$ multiplet and 17.5\,$\mu$s for the $^{5}$F$_{5}$ multiplet.
At 290\,K we obtain lifetimes of 219\,$\mu$s
for the $^{5}$S$_{2}$ multiplet and 8.5\,$\mu$s for the $^{5}$F$_{5}$ multiplet. 

The temperature dependence of the $^5$F$_5$ lifetime was measured over the range 10 K to 290 K. Figure \ref{fig:5F5decay} shows the inferred decay rate (in s$^{-1}$) as a function of temperature, which increases by a factor of approximately two over the temperature range measured. The experimental decay rate, $W(T)$, is the sum of  radiative, $W_\text{R}$, and non-radiative,  $W_\text{NR}$, contributions such that:
\begin{equation}
\label{eq:decay}
 W(T)= W_\text{R}(T)+ W_\text{NR}(T)
\end{equation}
If we assume that the radiative contribution does not depend on temperature, we can approximate the behaviour of the total decay rate by considering non-radiative relaxation from $^{5}$F$_{5}$ to the next-lowest energy multiplet, $^{5}$I$_{4}$. The energy separation between these multiplets is approximately 2000 cm$^{-1}$. The thermal dependence of the non-radiative decay rate arises from stimulated phonon emission at elevated temperatures which we write as:
\begin{equation}
\label{eq:NRdecay}
 W_\text{NR}(T)= W_\text{NR}(0)(n+1)^p
\end{equation}
where  $W_\text{NR}(0)$ is the non-radiative decay rate at 0\,K, $n=[\exp(\hbar \omega/k_\text{B}T)-1]^{-1}$ are Bose-Einstein factors for band phonon modes having an energy of $\hbar \omega$ with $k_\text{B}$ the Boltzmann constant, and $p$ is the order of the process.
This formulation makes the assumption that phonon absorption can be neglected and that the phonon density of states can be treated as if a single effective phonon energy is responsible for the decay. The maximum phonon energy in K$_{2}$YF$_{5}$ is approximately 480 cm$^{-1}$ \cite{tuyen2020k2yf5}, thus a minimum of four phonons are required to conserve energy.

A good fit to the data can be achieved assuming a five phonon decay process with $\hbar \omega=400$ cm$^{-1}$, $p=5$,  $W_\text{R}=10,000$\,s$^{-1}$, $W_\text{NR}(0) =48,000$\,s$^{-1}$. It is notable that the effective phonon energy used here is close to the maximum intensity band in the Raman scattering spectrum of K$_{2}$YF$_{5}$ \cite{tuyen2020k2yf5}.

As a check on our analysis, we can compare our derived radiative and non-radiative rates to measurements and calculations for Ho$^{3+}$ in LiYF$_4$ crystals  
\cite{Yuan2025}. In that study, the room-temperature lifetime of the $^{5}$F$_{5}$ multiplet was  measured to be 35.5\,$\mu$s and the Judd-Ofelt analysis gives
$W_\text{R} =  1770$\, s$^{-1}$. The deduced radiative and non-radiative decay rates for the $^{5}$F$_{5}$ multiplet of  Ho$^{3+}$  are much larger in K$_{2}$YF$_{5}$ than in  LiYF$_4$, reflecting the lower symmetry of K$_{2}$YF$_{5}$.

\clearpage

\subsection{Upconversion fluorescence in K$_{2}$YF$_{5}$ microparticles co-doped with Ho$^{3+}$ and Yb$^{3+}$}

\begin{figure}[b!]
\centering
 \includegraphics[width=0.5\textwidth]{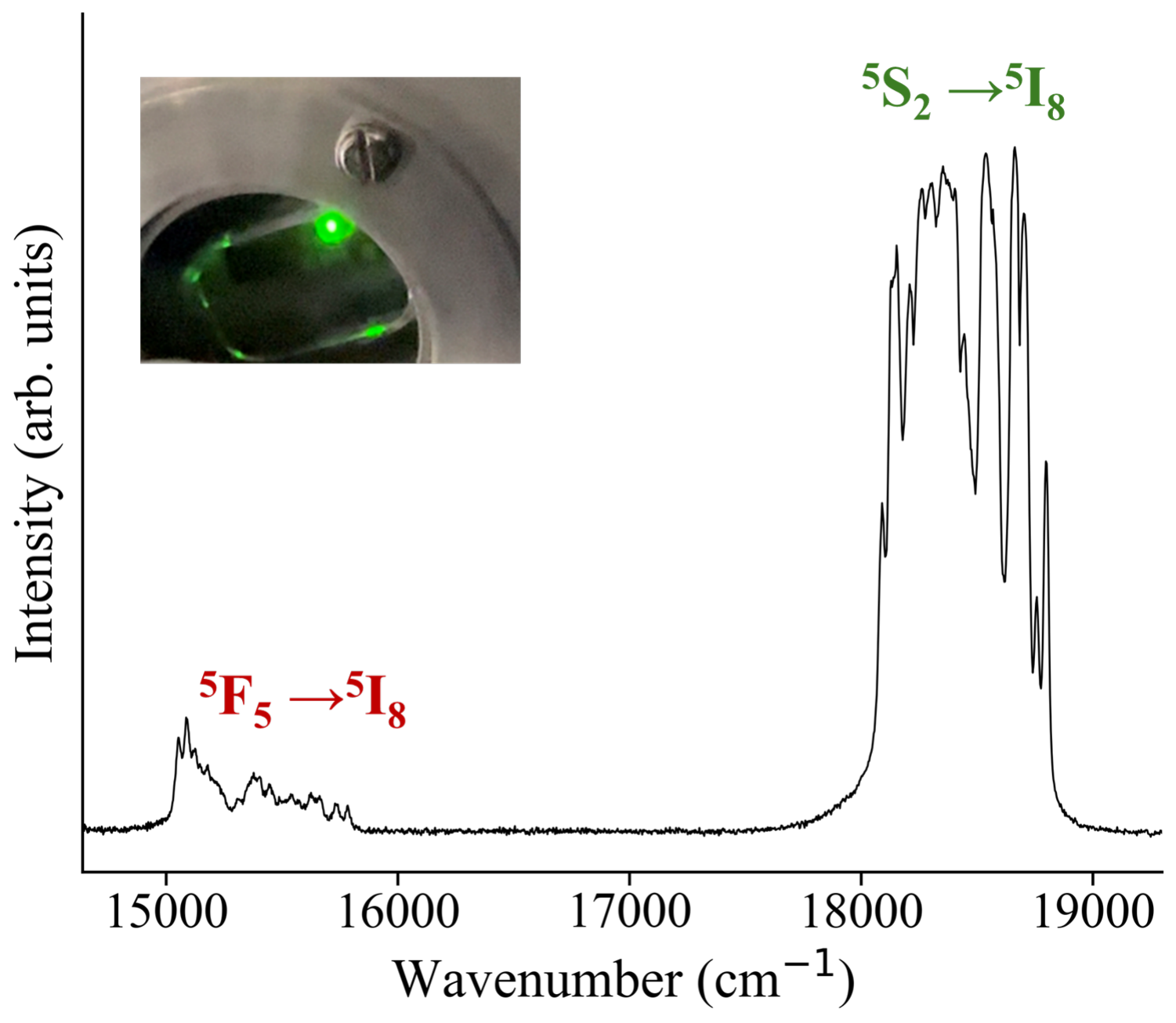} 
\caption{ \label{fig:upconversion}
The 10~K upconversion fluorescence spectrum of K$_{2}$YF$_{5}$:20\%Yb$^{3+}$:2\%Ho$^{3+}$ for excitation at 980 nm. Inset: bright green upconverted fluorescence from a microparticle pellet.
}
\end{figure}

\begin{figure}[b!]
\centering
 \includegraphics[width=1\textwidth]{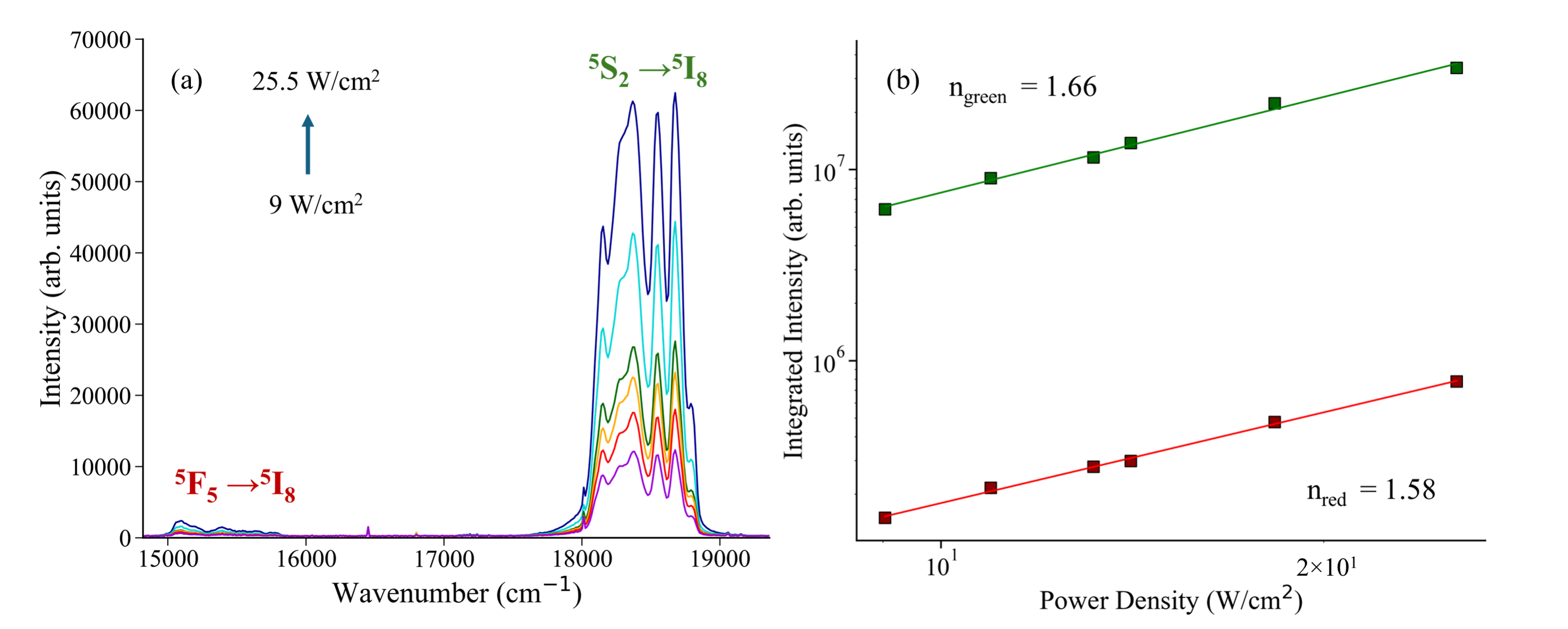} 
\caption{ \label{fig:powerdep}
(a) 10~K upconversion fluorescence spectra as a function of excitation power density between 9 W/cm$^2$ to 25.5 W/cm$^2$.
(b) Integrated upconversion intensity as a function of excitation power density. The solid lines are fits to equation \ref{eq:up}.}
\end{figure}

There is considerable current interest in optical imaging and contactless thermometry using micro- or nano-particles. Therefore, we have made a preliminary investigation of 20\%Yb$^{3+}$ co-doped K$_{2}$YF$_{5}$:2\%Ho$^{3+}$. Figure \ref{fig:upconversion} shows the low resolution upconversion fluorescence spectrum measured at 10 K, for excitation at 980 nm with a multi-quantum well InGaAs laser diode. As expected, strong holmium fluorescence is recorded, dominated by the $^{5}$S$_{2} {\rightarrow} ^{5}$I$_{8}$ transitions and therefore the pellet is observed to glow brightly green as can be seen in the inset to Figure \ref{fig:upconversion}. The order of the excitation process can be determined from the dependence of the upconversion fluorescence upon the excitation power density as \cite{Pollnau2000}:
\begin{equation}
\label{eq:up}
    I \propto (P_{\text{pump}})^n
\end{equation}
where $I$ is the integrated emission intensity, $P_{\rm pump}$ is the excitation power, and $n$ is the number of the photons involved in the process.
This is illustrated in Figure \ref{fig:powerdep} where the excitation power density was varied from 9 W/cm$^2$ to 25.5 W/cm$^2$. It is clear that for both the $^{5}$S$_{2}$ and $^{5}$F$_{5}$ multiplet upconversion fluorescence, a minimum of two photons are required \cite{Liu2013, Liu2023}.

\section{Conclusions}
K$_2$YF$_5$:Ho$^{3+}$ microparticles having lengths of roughly 10 $\mu$m were successfully prepared using a hydrothermal technique. From a combination of absorption and fluorescence measurements, a total of 72 crystal-field levels of Ho$^{3+}$ were determined. Partially-resolved measurements of the hyperfine structure provided additional input for crystal-field modelling. The temperature dependence of the fluorescent lifetime of the $^5$F$_5$ multiplet was modelled using a multi-phonon process.  Infra-red to visible upconversion in microparticles co-doped with Ho$^{3+}$ and Yb$^{3+}$ was also obtained.

Magnetic splitting measurements, such as EPR,  on single crystals would allow more accurate crystal-field modelling, not dependent on using results from Er$^{3+}$. Our calculations suggest that hyperfine structure should be readily resolvable in a high-quality bulk crystal of low Ho$^{3+}$ concentration.

\section{Acknowledgements}

The technical support of Mr.\ Stephen Hemmingson, Mr.\ Graeme MacDonald, Dr.\ Jamin Martin, Mr.\ Matthew Pannell, 
Dr.\  Matthew Polson, and Mr.\ Robert Thirkettle is gratefully acknowledged.
P.C.\ is grateful for the award of a Ph.D.\ scholarship from the Development and Promotion of Science and Technology Talents Project of the Royal Thai Government.

\printcredits

\clearpage


\bibliographystyle{elsarticle-num}
\bibliography{HoK2YF5_references.bib}

\begin{thebibliography}{10}
\expandafter\ifx\csname url\endcsname\relax
  \def\url#1{\texttt{#1}}\fi
\expandafter\ifx\csname urlprefix\endcsname\relax\def\urlprefix{URL }\fi
\expandafter\ifx\csname href\endcsname\relax
  \def\href#1#2{#2} \def\path#1{#1}\fi

\bibitem{Dong2015}
H.~Dong, S.~R. Du, X.~Y. Zheng, G.~M. Lyu, L.~D. Sun, L.~D. Li, P.~Z. Zhang,
  C.~Zhang, C.~H. Yan, Lanthanide nanoparticles: from design toward bioimaging
  and therapy, Chemical Reviews 115 (2015) 10725--10815.

\bibitem{Bouzigues2011}
C.~Bouzigues, T.~Gacoin, A.~Alexandrou, Biological applications of rare-earth
  based nanoparticles, ACS Nano 5 (2011) 8488--8505.

\bibitem{Himmelsto}
S.~F. Himmelsto{\ss}, T.~Hirsch, A critical comparison of lanthanide based
  upconversion nanoparticles to fluorescent proteins, semiconductor quantum
  dots, and carbon dots for use in optical sensing and imaging, Methods and
  Applications in Fluorescence 7 (2019) 022002.

\bibitem{Pratik2021}
P.~S. Solanki, S.~Balabhadra, M.~F. Reid, V.~B. Golovko, J.-P.~R. Wells,
  Upconversion thermometry using {Er$^{3+}$/Yb$^{3+}$} co-doped
  {KY$_{3}$F$_{10}$} nanoparticles, ACS Applied Nano Materials 117 (2021)
  111114.

\bibitem{Thiel2011}
C.~W. Thiel, T.~B{\"o}ttger, R.~L. Cone, Rare-earth-doped materials for
  applications in quantum information storage and signal processing, Journal of
  Luminescence 131 (2011) 353--361.

\bibitem{Martin2023}
J.~L.~B. Martin, L.~F. Williams, M.~F. Reid, J.-P.~R. Wells, Growth and
  spectroscopy of lanthanide doped {Y$_2$SiO$_5$} microcrystals for quantum
  information processing, Optical Materials 142 (2023) 114093.

\bibitem{karbowiak2012energy}
M.~Karbowiak, P.~Gnutek, C.~Rudowicz, Energy levels and crystal field
  parameters for {Nd$^{3+}$} ions in {BaY$_2$F$_8$}, {LiKYF$_5$}, and
  {K$_2$YF$_5$} single crystals, Spectrochimica Acta Part A: Molecular and
  Biomolecular Spectroscopy 87 (2012) 46--60.

\bibitem{zverev2011electron}
D.~G. Zverev, H.~Vrielinck, E.~Goovaerts, F.~Callens, Electron paramagnetic
  resonance study of rare-earth related centres in {K$_2$YF$_5$:Tb$^{3+}$}
  thermoluminescence phosphors, Optical Materials 33 (2011) 865--871.

\bibitem{loiko2016up}
P.~A. Loiko, N.~M. Khaidukov, J.~M{\'e}ndez-Ramos, E.~V. Vilejshikova, N.~A.
  Skoptsov, K.~V. Yumashev, Up-and down-conversion emissions from {Er$^{3+}$}
  doped {K$_2$YF$_5$} and {K$_2$YbF$_5$} crystals, Journal of Luminescence 170
  (2016) 1--7.

\bibitem{Mujaji1992}
M.~Mujaji, G.~D. Jones, R.~W.~G. Syme, Polarization study and crystal-field
  analysis of the laser-selective excitation spectra of {Ho$^{3+}$} ions in
  {CaF$_2$} and {SrF$_2$} crystals, Physical Review B 46 (1992) 14398.

\bibitem{Wells2004}
J.-P.~R. Wells, G.~D. Jones, M.~F. Reid, M.~N. Popova, E.~P. Chukalina,
  Hyperfine patterns of infrared absorption lines of {Ho$^{3+}$} {C$_{4v}$}
  centres in {CaF$_2$}, Molecular Physics 102 (2004) 1367--1376.

\bibitem{Popova2017}
M.~N. Popova, K.~N. Boldyrev, High-resolution spectra of {LiYF$_4$}:{Ho$^{3+}$}
  in a magnetic field, Optical Materials 63 (2017) 101--104.

\bibitem{Smith2023}
K.~M. Smith, M.~F. Reid, J.-P.~R. Wells, Zeeman-hyperfine measurements of a
  pseudo-degenerate quadruplet in {CaF$_{2}$:Ho$^{3+}$}, Optics and
  Spectroscopy 131 (2023) 460--465.

\bibitem{Mothkuri2021}
S.~Mothkuri, M.~F. Reid, J.-P.~R. Wells, E.~Lafitte-Houssat, P.~Goldner,
  A.~Ferrier, Electron-nuclear interactions as a test of crystal-field
  parameters for low symmetry systems: {Z}eeman hyperfine spectroscopy of
  {Ho$^{3+}$} -doped {Y$_{2}$SiO$_5$}, Physical Review B 103 (2021) 104109.

\bibitem{Mothkuri2024}
S.~Mothkuri, M.~F. Reid, J.-P.~R. Wells, E.~Lafitte-Houssat, A.~Ferrier,
  P.~Goldner, Laser site-selective spectroscopy and magnetic hyperfine
  splittings of {Ho$^{3+}$} doped {Y$_2$SiO$_5$}, Journal of Luminescence 275
  (2024) 120705.

\bibitem{Bian2019}
X.~Bian, Q.~Shi, L.~Wang, Y.~Tian, B.~Xu, Z.~K. Mamytbekov, P.~Huang, et~al.,
  Near-infrared luminescence and energy transfer mechanism in
  {K$_2$YF$_5$}:{Nd$^{3+}$,Yb$^{3+}$}, Materials Research Bulletin 110 (2019)
  102--106.

\bibitem{wybourne1965spectroscopic}
B.~G. Wybourne, Spectroscopic Properties of Rare Earths, {Interscience
  Publishers}, 1965.

\bibitem{Carnall1989}
W.~T. Carnall, G.~L. Goodman, K.~Rajnak, R.~S. Rana, A systematic analysis of
  the spectra of the lanthanides doped into single crystal {LaF$_3$}, Journal
  of Chemical Physics 90 (1989) 3443--3457.

\bibitem{reid2016theory}
M.~F. Reid, Theory of rare-earth electronic structure and spectroscopy, in:
  Handbook on the Physics and Chemistry of Rare Earths, Vol.~50, Elsevier,
  2016, pp. 47--64.

\bibitem{Solanki2024}
P.~S. Solanki, M.~F. Reid, J.-P.~R. Wells, Spectroscopy and crystal-field
  analysis of low-symmetry {Er$^{3+}$} centres in {K$_2$YF$_5$} microparticles,
  Optical Materials: X 24 (2024) 100356.

\bibitem{Gomes1996}
L.~Gomes, L.~C. Courrol, L.~V.~G. Tarelho, I.~M. Ranieri, Cross-relaxation
  process between +3 rare earth ions in {LiYF$_{4}$} crystals, Physical Review
  B 54 (1996) 3825.

\bibitem{Dieke}
G.~H. Dieke, Spectra and Energy Levels of Rare-Earth Ions in Crystals,
  {Interscience Publishers}, 1968.

\bibitem{horvath2019}
S.~P. Horvath, J.~V. Rakonjac, Y.~H. Chen, J.~J. Longdell, P.~Goldner, J.-P.~R.
  Wells, M.~F. Reid, Extending phenomenological crystal-field methods to
  {C$_{1}$} point group symmetry: characterization of the optically excited
  hyperfine structure of {$^{165}$Er$^{3+}$:Y$_2$SiO$_5$}, Physical Review
  Letters 123 (2019) 057401.

\bibitem{tuyen2020k2yf5}
V.~P. Tuyen, V.~X. Quang, N.~M. Khaidukov, L.~D. Thanh, N.~X. Ca, N.~Van~Hao,
  N.~Van~Nghia, P.~Van~Do, {K$_2$YF$_5$:Tb$^{3+}$} single crystal: An in-depth
  study of spectroscopic properties, energy transfer and quantum cutting,
  Optical Materials 106 (2020) 109939.

\bibitem{Yuan2025}
W.~Yuan, R.~Dai, Z.~Zhang, X.~Lin, H.~Xu, Z.~Cai, Spectral analysis and
  emission properties of {Ho$^{3+}$}: {YLF} in the visible-near infrared band,
  Journal of Luminescence 277 (2025) 120915.

\bibitem{Pollnau2000}
M.~Pollnau, D.~R. Gamelin, S.~R. L{\"u}thi, H.~U. G{\"u}del, M.~P. Hehlen,
  Power dependence of upconversion luminescence in lanthanide and
  transition-metal-ion systems, Physical Review B 61 (2000) 3337.

\bibitem{Liu2013}
W.~Liu, J.~Sun, X.~Li, J.~Zhang, Y.~Tian, S.~Fu, H.~Zhong, T.~Liu, L.~Cheng,
  H.~Zhong, et~al., Laser induced thermal effect on upconversion luminescence
  and temperature-dependent upconversion mechanism in
  {Ho$^{3+}$/Yb$^{3+}$}-codoped {Gd$_2$(WO$_4$)$_3$} phosphor, Optical
  Materials 35 (2013) 1487--1492.

\bibitem{Liu2023}
R.~Liu, Z.~Lei, R.~Sun, S.~Su, Y.~Zou, L.~Sun, J.~Lu, C.~Hu, B.~Teng, S.~Sun,
  et~al., Broad-scope dual-mode modulation thermometry based on up-conversion
  phosphor {GaNbO$_4$}:{Yb$^{3+}$/Er$^{3+}$}, Ceramics International 49 (2023)
  11829--11836.

\end{thebibliography}

\end{document}